\documentclass[preprint,showpacs,amsmath,amssymb,aps,prd,nofootinbib]{revtex4}
\usepackage{epsfig}
\begin{document}

\def\bea{\begin{eqnarray}}
\def\eea{\end{eqnarray}}
\def\nn{\nonumber}
\newcommand{\snu}{\tilde \nu}
\newcommand{\sll}{\tilde{l}}
\newcommand{\asnu}{\bar{\tilde \nu}}
\newcommand{\stau}{\tilde \tau}
\newcommand{\dmsnu}{{\mbox{$\Delta m_{\tilde \nu}$}}}
\newcommand{\mt}{{\mbox{$\tilde m$}}}

\renewcommand\epsilon{\varepsilon}
\def\be{\begin{eqnarray}}
\def\ee{\end{eqnarray}}
\def\lla{\left\langle}
\def\rra{\right\rangle}
\def\za{\alpha}
\def\zb{\beta}
\def\lsim{\mathrel{\raise.3ex\hbox{$<$\kern-.75em\lower1ex\hbox{$\sim$}}} }
\def\gsim{\mathrel{\raise.3ex\hbox{$>$\kern-.75em\lower1ex\hbox{$\sim$}}} }
\newcommand{\Rbs}{\mbox{${{\scriptstyle \not}{\scriptscriptstyle R}}$}}

\draft


\title{Extended double seesaw model for neutrino mass spectrum \\
       and low scale leptogenesis}

\textwidth 16.5cm
\textheight 24.5cm

\thispagestyle{empty}
\author{ Sin Kyu Kang\footnote{ E-mail : skkang1@sogang.ac.kr}}
\affiliation{ Center for Quantum Spacetime and Department of Physics, Sogang University,
       Seoul 121-742, Korea }
\author{C. S. Kim\footnote{ E-mail : cskim@yonsei.ac.kr}}
\affiliation{ Department of Physics, Yonsei University, Seoul
120-749, Korea}

\begin{abstract}
\noindent We consider a variant of seesaw mechanism by introducing
extra singlet neutrinos, with which we show how the low scale
leptogenesis is realized  without imposing the tiny mass splitting
between two heavy Majorana neutrinos required in the resonant
leptogenesis. Thus, we can avoid the so-called gravitino problem
when our scenario is supersymmetrized. We show that an
introduction of the new singlet fermion leads to a new
contribution which can enhance the lepton asymmetry for
certain range of parameter space.  We also examine how both
the light neutrino mass spectrum and relatively light sterile
neutrinos of order a few 100 MeV can be achieved without being in
conflict with the constraints on the mixing between the active and
sterile neutrinos.

\end{abstract}
 \maketitle \thispagestyle{empty}
%
\textwidth 18cm
\textheight 24.5cm

There exist quite many motivations for postulating the existence of weak interaction singlets
such as singlet neutrinos which may arise in most extensions of the standard model (SM);
It is by now well known that the smallness of neutrino masses can be understood by introducing
heavy singlet Majorana neutrinos via the so-called seesaw mechanism \cite{seesaw}.
It was also noticed that
sterile neutrinos  could be a good viable candidate for cosmological dark matter \cite{stdm}.
The controversial result from the LSND experiment \cite{lsnd} seems to indicate that the complete
picture of neutrino sector has need additional neutrinos beyond the three ordinary ones,
which must be sterile.
The sterile neutrino states can mix with the active neutrinos and such admixtures contribute
to various processes which are forbidden in the SM, and affect the interpretations of cosmological
and astrophysical observations. Thus, the masses of the sterile neutrinos and their mixing with
the active neutrinos are subject to various experimental bounds as well as cosmological and
astrophysical constraints.

In addition to the accomplishment of smallness for  neutrino masses, another virtue
of the seesaw mechanism is that it can be responsible for baryon asymmetry of our Universe
via so-called leptogenesis \cite{lepto}.
However, the relevant scale of the typical leptogenesis with
hierarchical heavy Majorana neutrino mass spectrum is larger than
$\sim 10^{9}$ GeV \cite{davidsn}, which makes it impossible to probe the seesaw
model at collider experiments in a foreseeable future. And thermal
leptogenesis working at such a high mass scale encounters
gravitino problem in the framework of supersymmetric SM as
well. Thus, a low scale seesaw is desirable to reconcile such
proplems.
We may emphasize that it is also quite nontrivial to naturally realize the
hierarchy of the neutrino mass spectrum in the seesaw model.

Resonant leptogenesis with  very tiny mass splitting ($(M_2-M_1)/(M_2+M_1)\sim
10^{-6)}$) of heavy Majorana neutrinos with $M_1\sim 1$ TeV
has been proposed as a successful scenario for  a low scale leptogenesis \cite{pilaft}.
However, such a very
tiny mass splitting  may appears somewhat unnatural due to the required severe fine-tuning.
In order to remedy
such problems mentioned above
in this letter, we consider a variant of the seesaw mechanism,
in which an equal number of gauge singlet neutrinos is introduced on
top of the heavy right-handed neutrinos. The model we consider looks
similar to the so-called double seesaw mechanism \cite{dss}, but is
obviously different from it because mass terms of the heavy
right-handed neutrinos, which are not present in the typical double
seesaw model, are here introduced \cite{flipped}. Such additional
sterile neutrinos can be naturally incorporated into the superstring
$E_6$ \cite{dss} and/or flipped SU(5) GUT \cite{flipped}. Unlike to
the typical double seesaw model, as will be shown later, our model
permit both tiny neutrino masses and relatively light sterile
neutrinos of order MeV which may play important roles in cosmology
and astrophysics. It can also accommodate very tiny mixing between the
active and sterile neutrinos demanded from the cosmological and
astrophysical observations. Note that the effects of the heavy
right-handed neutrinos are cancelled in our seesaw model for the
tiny neutrino masses, whereas they may play an important role in
generating MeV sterile neutrinos. In particular, we will show that
a low scale thermal leptogenesis  can be naturally
achieved without encountering the problems described above. And we will examine how
both the light neutrino mass spectrum and the sterile neutrino mass of order a few
100 MeV can be obtained without being in conflict with the constraints on the
mixing between the active and sterile neutrinos.
In this scenario, the mass terms of the newly introduced singlet neutrinos
are responsible for the hierarchy of the light neutrino mass spectrum. Thus, the
structure of the light neutrino mass matrix may be determined by the
mechanism of screening of the Dirac flavor structure \cite{screen}
in our model.

\vspace{0.3cm}
The Lagrangian we propose is given in the charged
lepton basis as
\be
{\cal L}=M_{R_i}N_i^T N_i+Y_{D_{ij}} \bar{\nu}_i
\phi  N_j+ Y_{S_{ij}} \bar{N}_i\Psi S_j -\mu_{ij} S_i^T S_j +h.c.~,
\ee
where $\nu_i,N_i,S_i$ stand for SU(2)$_L$ doublet, right-handed
singlet and newly introduced singlet neutrinos, respectively. And
$\phi$ and $\Psi$ denote the SU(2)$_L$ doublet and singlet Higgs
sectors. The neutrino mass matrix  after the scalar fields get
vacuum expectation values becomes
\be
M_{\nu}=\left(\begin{array}{ccc}
 0 & m_{D_{ij}} & 0 \\
 m_{D_{ij}} & M_{R_{ii}} & M_{ij} \\
 0 & M_{ij} & -\mu_{ij} \end{array}\right), \label{massmatrix}
\ee
where $m_{D_{ij}}=Y_{D_{ij}}<\phi>, M_{ij}=Y_{S_{ij}}<\Psi>$.
Here we assume that $M_R>M \gg \mu, m_{D}$. After integrating out
the right-handed heavy neutrino sector $N_R$ in the above
Lagrangian, we obtain the following effective lagrangian,
\be
-{\cal L}_{eff}&=& \frac{({m_{D}^2})_{ij}}{4M_R}\nu_i^T \nu_j+
 \frac{m_{D_{ik}}M_{kj}}{4M_R}(\bar{\nu}_i S_j+\bar{S_i}\nu_j)\nonumber \\
& & +\frac{M^2_{ij}}{4M_R}S_i^TS_j
 + \mu_{ij} S_i^T S_j.
\ee
After block diagonalization of the effective mass terms in ${\cal
L}_{eff}$, the light neutrino mass matrix and mixing between the
active and sterile neutrinos are given by
\be
m_{\nu} &\simeq & \frac{1}{2}\frac{m_D}{M}~\mu~ \left(\frac{m_D}{M}\right)^T,
                  \label{dsw}\\
\tan2\theta_s &=& \frac{2m_DM}{M^2+4\mu M_R-m_D^2}~, \label{mixing}
\ee
where we omitted the indices of the mass matrices, $m_D,M,M_R$ and $\mu$
for simplicity.

We note that the term $m_D^2/M_R$ corresponding to
typical seesaw (type I) mass is cancelled out.
On the other hand, the sterile neutrino mass is approximately
given as
\begin{eqnarray}
m_s \simeq \mu + \frac{M^2}{4M_R}. \label{sterile}
\end{eqnarray}
Depending on the relative sizes among $M,M_R,\mu$, the mixing angles
$\theta_s$ and the sterile neutrino mass $m_s$ are approximately
given by
\be
\tan2\theta_s \simeq \sin2\theta_s &\simeq &\left\{
\begin{array}{cc}
 \frac{2m_D}{M} & ~~~(\mbox{for}~~M^2>4\mu M_R~: ~~~~\mbox{\bf Case A}), \\
  \frac{m_D}{M} &  ~~~(\mbox{for}~~M^2\simeq 4\mu M_R~: ~~~~ \mbox{\bf Case B}),\\
  \frac{m_DM}{2\mu M_R} & ~~~(\mbox{for}~~M^2 < 4\mu M_R~: ~~~~ \mbox{\bf Case C}),
  \end{array}
  \right. \\
{\rm and}~~~~m_s &\simeq &\left\{
\begin{array}{cc}
 \frac{M^2}{4M_R} & ~~~~(\mbox{\bf Case A}), \\
  2\mu &   ~~~~(\mbox{\bf Case B}),\\
  \mu &  ~~~~(\mbox{\bf Case C}).
  \end{array}
  \right.  \label{sterile2}
\ee
We can see from Eq. (\ref{sterile2}) that for $M^2\leq 4\mu M_R$, the size of $\mu$ is mainly
responsible for the value of $m_s$.
We also notice that the value of the mixing angle $\theta_s$
is suppressed by the scale of $M$ or $M_R$. Thus, very small mixing angle
$\theta_s$ can be naturally achieved in our seesaw mechanism.
As expected, for {\bf Case A} and {\bf Case B}, the mixing angle $\theta_s$ constrained
by the cosmological and astrophysical observation as well as
laboratory experiments leads to constraints on the size of the
ratio $m_{\nu}/\mu$ through the Eq. (\ref{dsw}) .

\vspace{0.3cm}
The existence of a relatively light sterile neutrino has
nontrivial observable consequences for cosmology and astrophysics,
so that the masses of sterile neutrinos and their mixing with the
active ones must be subject to the cosmological and astrophysical
bounds \cite{smirnovC}. There are also some laboratory bounds
which are typically much weaker than the astrophysical and
cosmological ones. Those bounds turn out to be useful in the case
that the cosmological and astrophysical limits become
inapplicable. In the light of laboratory experimental as well as
cosmological and astrophysical observations, there exist two
interesting mass ranges of the sterile neutrinos, one is of order
keV, and the other of order MeV.
For those mass ranges of  sterile
neutrinos, their mixing with the active ones are constrained by
various cosmological and astrophysical observations as described
below.

It has been shown that a sterile neutrino with the mass in
keV range appears to be a viable ``warm" dark matter candidate
\cite{warm}. The small mixing angle ($\sin\theta \sim
10^{-6}-10^{-4}$) between sterile and active neutrinos ensures that
sterile neutrinos were never in thermal equilibrium in the early
Universe and this allows their abundance to be smaller than the
predictions in thermal equilibrium. Moreover, a sterile neutrino with these
parameters is important for the physics of supernova, which can
explain the pulsa kick velocities \cite{pulsa}. In addition, there
are some bounds on the mass of sterile neutrino from the
possibility to observe sterile neutrino radiative decays from
X-ray observations and Lyman $\alpha-$forest observations of order
of a few keV.

On the other hand,  there exists high mass region ($m_s \gtrsim 100$ MeV) of the
sterile neutrinos, which is restricted by the CMB bound, meson decays and SN1987A cooling.
For this mass range of sterile neutrino,
the mixing between the active and sterile neutrinos is constrained to be approximately
$\sin^2\theta_s  \lesssim 10^{-9}$ \cite{smirnovC}.
Such a high mass region may be very interesting, because
induced contributions to the active neutrino
mass matrix due to mixing between the active  and sterile neutrinos  can be dominant
and they can be responsible for peculiar properties of the
lepton mixing such as tri-bimaximal mixing and neutrino mass spectrum \cite{smirnovC}.
In addition, sterile neutrinos with mass 1-100 MeV can be a dark matter candidate for the
explanation of the excess flux of 511 keV photons if
$\sin^2 2\theta_s \lesssim 10^{-17}$ \cite{sterile511}.
The detailed constraints coming from current astrophysical,
cosmological and laboratory observations are presented in the figures of Ref. \cite{smirnovC}.
In this letter, we will focus on sterile neutrinos with masses in the range of MeV.
Similarly, we can realize keV sterile neutrinos but at the price of naturalness for the magnitudes
of some parameters.

\vspace{0.3cm}
First of all, let us consider how low scale leptogenesis can be
successfully achieved in our scenario without severe fine-tuning
such as very tiny mass splitting between two heavy Majorana neutrinos.
In our scenario, the successful leptogenesis can be achieved by the
decay of the lightest right-handed Majorana neutrino before the
scalar fields get vacuum expectation values. As will be discussed, when
our scenario is supersymmetrized, we can escape the so-called
gravitino problem  because  rather light right-handed
Majorana neutrino masses are possible in our scenario. In
particular, there is a new contribution to the lepton asymmetry which
is mediated by the extra singlet neutrinos.

Without loss of generality, we can rotate and rephase the fields to
make the mass matrices $M_{R_{ij}}$ and $\mu_{ij}$ real and
diagonal. In this basis, the elements of $Y_D$ and $Y_S$ are in
general  complex. The lepton number asymmetry required for
baryogenesis is given by
\begin{eqnarray}
\epsilon_{1} &=& -\sum_i\left[\frac{\Gamma(N_1
\to \bar{l_i}\bar{H}_u) - \Gamma(N_1 \to l_iH_u) }{\Gamma_{\rm
tot}(N_1)}\right] ,
\end{eqnarray}
where $N_1$ is the lightest right-handed neutrino and $\Gamma_{\rm tot}(N_1)$
is the total decay rate.
In our scenario, the introduction of the new singlet fermion leads
to a new contribution to $\epsilon_{1}$ which for certain range of
parameters can enhance the effect. We show that as a result of this
enhancement, even for $M_1$ as low as a few TeV, we can have
successful leptogenesis while keeping tiny masses for neutrinos.
Therefore, thermal production of $N_1$  does not need too high
reheating temperature and the Universe would not encounter the gravitino
overproduction \cite{Kawasaki}.

\begin{figure}[htbp]
 \centering
  \epsfig{figure=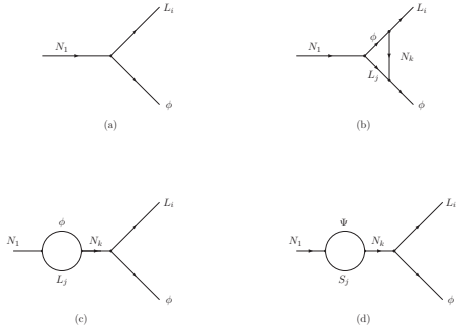, width=6cm}
  \caption{Diagrams contributing to lepton asymmetry.
     }
  \label{fig1}
\end{figure}

Fig. 1 shows the structure of the diagrams contributing to
$\epsilon_{1}$. In addition to the diagrams of the standard
leptogensis scenario \cite{Covi}, there is a new contribution of the
diagram which corresponds to the self energy correction of the
vertex arisen due to the new Yukawa couplings with singlet neutrinos
and Higgs sectors.   Assuming that the masses of the Higgs sectors
and extra singlet neutrinos are much smaller compared to that of the
right-handed neutrino, to leading order, we have
\begin{equation}
  \label{eq:vv}
\Gamma_{\rm tot}(N_i)={(Y_\nu Y_\nu^\dagger+Y_sY_s^\dagger)_{ii}
\over 4\pi}M_{R_i}
\end{equation}
so that
\begin{equation}
\epsilon_1 = \frac{1}{8\pi} \sum_{k\ne 1} \left( [ g_V(x_k)+
g_S(x_k)]{\cal T}_{k1} + g_S(x_k){\cal S}_{k1}\right),
\end{equation}
where $g_V(x)=\sqrt{x}\{1-(1+x) {\rm ln}[(1+x)/x]\}$,
$g_S(x)=\sqrt{x_k}/(1-x_k) $ with $x_k=M_{R_k}^2/M_{R_1}^2$ for
$k\ne 1$,
\begin{equation}
  \label{eq:vv}
{\cal T}_{k1}={{\rm Im}[(Y_\nu Y_\nu^\dagger)_{k1}^2]
 \over (Y_\nu Y_\nu^\dagger +Y_s Y_s^\dagger)_{11}}
\end{equation}
and
\begin{equation}
{\cal S}_{k1}={{\rm Im}[(Y_\nu Y_\nu^\dagger)_{k1}(Y_s^\dagger
Y_s)_{1k}]
 \over (Y_\nu Y_\nu^\dagger +Y_s Y_s^\dagger)_{11}}.
\end{equation}
Notice that the term proportional to ${\cal S}_{k1}$ comes from the
interference of the tree-level diagram with diagram (d).

For $x \gg 1$, the contributions of the self-energy diagrams become
negligibly small, and thus the asymmetry can be approximately
written as \cite{DI},
\begin{eqnarray}
 \epsilon_1\simeq -\frac{3M_{R_1}}{16\pi v^2}
 \frac{Im[(Y^{\ast}_{\nu}m_{\nu}Y^{\dagger}_{\nu})_{11}]}
      {(Y_{\nu}Y^{\dagger}_{\nu}+Y_sY^{\dagger}_s)_{11}}, \label{epb}
 \end{eqnarray}
 and  thus  bounded as
 \begin{eqnarray}
 |\epsilon_1 |< \frac{3}{16\pi}\frac{M_{R_1}}{v^2}(m_{\nu_3}-m_{\nu_1}), \label{bound1}
 \end{eqnarray}
 where $m_{\nu_i}(i=1-3)$ stands for mass eigenvalues of three light neutrinos.
 For hierarchical neutrino mass spectrum, we can identify $m_{\nu_3}\simeq \sqrt{\Delta m^2_{atm}}$
 and then it is required that $M_{R_1}\geq 2\times 10^{9}$ GeV to achieve successful leptogenesis.

 On the other hand, for $x\simeq 1$, the vertex contribution to
 $\epsilon_1$ is much smaller than the contribution of the self-energy diagrams and
 the asymmetry $\epsilon$ is resonantly enhanced.
 To see how much the new contribution can be important in this case, for simplicity,
 we consider a particular situation where $M_{R_1}\simeq M_{R_2} < M_{R_3}$,
 so that the effect of $N_3$ is negligibly small.
 In this case, the asymmetry 
 can be written as
 \begin{eqnarray}
 \epsilon_1 & \simeq & -\frac{1}{16\pi}\left[
           \frac{M_{R_2}}{v^2}\frac{Im[(Y^{\ast}_{\nu}m_{\nu}Y^{\dagger}_{\nu})_{11}]}
           {(Y_{\nu}Y^{\dagger}_{\nu}+Y_sY^{\dagger}_s)_{11}}
           \right.
           \nonumber \\
    & & \left.  +\frac{\sum_{k\ne 1} Im[(Y_{\nu}Y^{\dagger}_{\nu})_{k1}(Y_sY_s^{\dagger})_{1k}]}
                   {(Y_{\nu}Y^{\dagger}_{\nu}+Y_sY^{\dagger}_s)_{11}}\right]R~,
                   \label{epsilon2}
 \end{eqnarray}
where $R$ is a resonance factor defined by $R \equiv
|M_{R_1}|/(|M_{R_2}|-|M_{R_1}|)$. For successful leptogenesis, the
size of the denominator of $\epsilon_1$ should be constrained by the
out-of-equilibrium condition, $\Gamma_{N_1} < H|_{T=M_{R_1}}$ with
the Hubble expansion rate $H$, from which the corresponding upper
bound on the couplings $(Y_s)_{1i}$ reads
\begin{eqnarray}
\sqrt{\sum_i|(Y_s)_{1i}|^2}<3\times
10^{-4}\sqrt{M_{R_1}/10^9(\mbox{GeV})}.
\end{eqnarray}
As shown in Eqs. (\ref{epb},\ref{bound1}), the first term of Eq.
(\ref{epsilon2}) is bounded as $(M_{R_2}/16\pi v^2)\sqrt{\Delta
m_{atm^2}}R$. If the first term of Eq. (\ref{epsilon2}) dominates
over the second one, $R\sim 10^{6-7}$ is required to achieve
TeV scale leptogenesis, which implies severe fine-tuning.
However, since the size of $(Y_s)_{2i}$ is not constrained by the
out-of-equilibrium condition, large value of $(Y_s)_{2i}$ is allowed
for which the second term of Eq. (\ref{epsilon2}) can dominate over
the first one and thus the size of $\epsilon_1$ can be enhanced.
For example, if we assume that $(Y_{\nu})_{2i}$ is aligned to
$(Y_s^{\ast})_{2i}$, {\it i.e.} $(Y_{s})_{2i}=\kappa
(Y_{\nu}^{\ast})_{2i}$ with constant $\kappa$, the upper limit of
the second term of Eq. (\ref{epsilon2}) is given in terms of
$\kappa$ by $|\kappa|^2 M_{R_2}\sqrt{\Delta m_{atm}^2}R/16\pi v^2$,
and then we can achieve the successful low scale leptogenesis by
taking rather large value of $\kappa$ instead of imposing very tiny
mass splitting between $M_{R_1}$ and $M_{R_2}$.

As one can estimate, the successful leptogenesis  can be achieved
for $M_{R_1}\sim$ a few TeV, provided that
$\kappa=(Y_s)_{2i}/(Y_{\nu})^{\ast}_{2i}\sim 10^{3}$ and
$M_{R_2}^2/M_{R_1}^2\sim 10$.
We {\it emphasize} that such
a requirement for the hierarchy between $Y_{\nu}$ and $Y_s$ is much
less severe than the required fine-tuning of the mass splitting
between two heavy Majorana neutrinos to achieve the successful
leptogenesis at low scale.

The generated B-L asymmetry is given by  $Y^{SM}_{B-L}=-\eta
\epsilon_1 Y^{eq}_{N_1}$, where $Y^{eq}_{N_1}$ is the number density
of the right-handed heavy neutrino at $T \gg M_{R_1}$ in thermal
equilibrium given by $Y^{eq}_{N_1}\simeq
\frac{45}{\pi^4}\frac{\zeta(3)}{g_{\ast}k_B} \frac{3}{4}$ with
Boltzmann constant $k_B$ and the effective number of degree of
freedom $g^{\ast}$. The efficient factor $\eta$ can be computed
through a set of coupled Boltzmann equations which take into account
processes that create or washout the asymmetry. To a good
approximation the efficiency factor depends on the effective
neutrino mass $\tilde{m}_1$ given in the presence of the new Yukawa
interactions with the coupling $Y_s$ by
\begin{eqnarray}
\tilde{m}_1=\frac{(Y_{\nu}Y^{\dagger}_{\nu}+Y_sY^{\dagger}_s)_{11}
}{M_{R_1}}v^2.
\end{eqnarray}
In our model, the new process of type $S\Psi \rightarrow l\phi$ will
contribute to wash-out of the produced B-L asymmetry. The process
occurs through virtual $N_{2,3}$ exchanges and the rate is
proportional to $M_{R_1}|Y_sY_{\nu}^{\ast}/M_{R_{2,3}}|^2$.
Effect of the wash-out can be easily estimated from the fact that it
looks similar to the case of the typical seesaw model if  $M_{R_1}$
is replaced with $M_{R_1}(Y_s/Y_{\nu})^2$. It turns out that the
wash-out factor for $(Y_s)_{1i}\sim (Y_\nu)_{1i}$,
$(Y_s)_{2i}/(Y_{\nu})_{2i}\sim 10^3$ and $M_{R_1}\sim 10^4$ GeV is
similar to the case of the typical seesaw model with  $M_{R_1}\sim
10^4$ GeV and $\tilde{m}_1\simeq 10^{-3}$ eV, and is estimated so
that $\epsilon_{1} \sim 10^{-6}$ can be enough to explain the baryon
asymmetry of the universe provided that the initial number of the
lightest right-handed neutrino is thermal \cite{Buchmuller}.
Detailed numerical calculation of the wash-out effect is beyond the
scope of this letter and will be presented elsewhere.

\vspace{0.3cm}
Now, let us examine how both light neutrino masses
of order 0.01 $\sim$ 0.1 eV and sterile neutrino masses of order
100 MeV can be simultaneously realized in our scenario
without being in conflict with the constraints on the mixing $\theta_s$.
Although the absolute values of three neutrinos are unknown, the largest and
the next largest neutrino masses are expected to be of order of
$\sqrt{\Delta m_{atm}^2} \simeq 0.05$ eV and $\sqrt{\Delta
m_{sol}^2} \simeq 0.01$ eV, provided that the mass spectrum of
neutrinos is hierarchical. There is also a bound on neutrino mass
coming from the WMAP observation, which is $m_{\nu}\lesssim 0.23$
eV assuming three almost degenerate neutrinos. Thus, it is
interesting to see how the neutrino mass of order of 0.01 $\sim$
0.1 eV can be obtained in our scenario.
We note that the low scale seesaw can be achieved by taking
$m_D$ to be 1-100 MeV which are of order of the
first and second charged lepton masses.
To begin with, for our numerical analysis,
we take $\sin^2\theta_s \simeq 10^{-9}$, which is allowed by
the constraints for the mass of sterile neutrinos $m_s \sim \mbox{a few}~~
100$ MeV as described before.
Then, we determine the values of the parameters so that the sterile neutrino mass
$m_s$ and the lightest heavy Majorana neutrino mass $M_{R_1}$ are
reached to be about 250 GeV and 1 TeV, respectively.

\vspace{0.3cm}
{\bf Case A} : For  $M^2 > 4\mu M_R$, $\sin^2\theta_s
\simeq (m_D/M)^2 $ and it follows from Eq. (\ref{dsw}) that
$m_{\nu_i} \simeq 0.5 \sin^2\theta_s \mu_i $. Then the value of the
light neutrino mass is determined to be $m_{\nu_{i}} \simeq$
0.01~(0.1) eV for $\mu_{i} \simeq 20~(200)$ MeV. For the given value
of $\sin^2\theta_s $, the size of $M_i$ should be $m_{D_i} \times
\sqrt{10^9}$, and thus the value of the lightest singlet neutrino
mass $M_1$ is determined to be 30 GeV for $m_{D_1}\sim 1$ MeV which
is order of electron mass. In this case, the lightest neutrino mass
$m_{\nu_1}$ depends on the size of $\mu_1$ which should be much less
than 20 MeV in the case of hierarchical light neutrino mass
spectrum. We also see from Eq. (\ref{sterile}) that $m_{s_1}\simeq
250$ MeV can be realized by taking $M_{R_1}\simeq 1$ TeV. As
presented in Eq. (\ref{sterile2}),  the size of $\mu$ does not
strongly affect the value of $m_{s}$ in this case.

As shown before, the successful
leptogenesis could be achieved for $M_{R_2}^2 \simeq 10\times
M_{R_1}^2$, and thus in order to obtain $m_{\nu_2}=0.01$ eV and
$m_{s_2}\simeq 250$ MeV, we require $M_{R_2}\simeq 3$ TeV and
$M_2\simeq 50$ GeV which is achieved for $m_{D_2}\simeq 1.7$ MeV.
However, for $\mu_i\simeq 200$ MeV which leads to $m_{\nu_i}\simeq 0.1$ eV,
the value of $m_{s_i}$ must be more than 400 MeV. Thus, to realize a few
100 MeV sterile neutrino we should include
the contribution of $\mu$ to $m_s$ as well.

\vspace{0.3cm}
{\bf Case B} : For $M^2 = 4\mu M_R$, $\tan2\theta_s
\simeq 2\sin\theta_s\simeq m_D/M$. Then, for $\sin^2\theta_s \simeq
10^{-9}$, the value of $(m_D/M)^2$ becomes $ 4\times 10^{-9}$, from
which the value of the light neutrino mass is determined to be
$m_{\nu_{i}} \simeq$ 0.01 (0.1) eV for $\mu_{i} \simeq 5$ ~(50) MeV.
The sizes of $M_i$ is given by $1.6\times 10^{4}~m_{D_i}$. In this
case, since $m_{s_i}\simeq 2 \mu_i$ as presented in Eq.
(\ref{sterile2}), one can achieve $m_s\simeq 100$ MeV which
corresponds to $m_{\nu_i}\simeq 0.1$, whereas the mass of $m_{s}$
corresponding to $m_{\nu_i}\simeq 0.01$ is determined to be at most
of order 10 MeV.
Thus, the case of hierarchical light neutrino spectrum disfavors
100 MeV sterile neutrinos for the first and second generations.

In particular, the size of $M_R$ is given by $M^2/(4\mu)\simeq 6\times 10^{7}m_{D}^2/\mu
\simeq 0.12m_{D}^2/m_{\nu}$, where we used Eqs. (\ref{dsw},\ref{mixing}) to obtain
the last relation.
For example, if we take $m_D\simeq 1$ MeV and $\nu\simeq 0.1$ eV, we obtain $M_R\simeq 1.2$ TeV.
Therefore, it is rather easy to achieve the low scale leptogenesis in consistent with
neutrino data as well as 100 MeV sterile neutrino in the case of the quasi-degenerate
light neutrino mass spectrum of order 0.1 eV.


\vspace{0.3cm}
{\bf Case C} :
For $ 4\mu M_R > M^2 $, $\tan2\theta_s \simeq 2\sin\theta_2\simeq m_DM/(2\mu M_R)$.
Combining this with Eq. (\ref{dsw}), we obtain
\begin{eqnarray}
\sin \theta_s = \frac{m_D^3}{8m_{\nu}MM_R}.
\end{eqnarray}
Then, the size of $MM_R$ should be $4\times 10^{5} ~(4\times 10^{11}) ~\mbox{GeV}^2$
for $\sin^2\theta_s \simeq 10^{-9}$ and $m_D=1 ~(100)$ MeV.
In this case, we have to  know the size of $M_i$ as well as $\mu_i$
in order to determine the light neutrino masses.
Recall that the value of $m_s$ strongly depends on that of $\mu$ as long as $4\mu M_R >> M^2$.

\vspace{0.3cm}
For smaller values of $\theta_s$, we note that larger value of
$\mu$ is demanded so as to achieve the required value of the light
neutrino mass.  In what follows, we present our numerical results for $\sin^2\theta_s=10^{-10}$.
Then,  $m_{\nu_i}
\simeq 0.5\times 10^{-10} \mu_i $, and the values of the light neutrino mass are
determined to be $m_{\nu_{i(=2,3)}} \simeq$ 0.01 ~(0.1) eV for
$\mu_{i} \simeq 200$ (~$2\times 10^3$) MeV in {\bf Case A} and for
$\mu_{i} \simeq 50$ ~(500) MeV in {\bf Case B}.
The size of $M_i$ should be $10^5\cdot m_{D_i}$ in {\bf Case A} and
$5\times 10^4\cdot m_{D_i}$ in {\bf Case B}.
Therefore, taking
$m_{D_1}=1$ MeV , we obtain
$M_{1}\simeq 100 $ GeV for {\bf Case A} and $M_{1}\simeq 50 $ GeV for {\bf Case B}.
In this case, we can obtain from Eq. (\ref{sterile}) that
$m_{s_1}\simeq 250$ MeV for $M_{R_1} \simeq 10 $ TeV  in {\bf
Case A}, which is desirable for the low scale leptogenesis.
For {\bf Case B}, 100 MeV sterile neutrinos can only be
achieved for the second generation. But, we see that the sterile
neutrino masses become GeV scale for the cases of $\mu_i=2\times10^3
(500)$ MeV in {\bf Case A (B)}.



\vspace{0.3cm}
In summary, we have considered a variant of seesaw mechanism by introducing extra singlet neutrinos
and investigated how the low scale leptogenesis is realized without imposing he tiny mass splitting
between two heavy Majorana neutrinos, which
makes us avoid the so-called gravitino problem when our scenario is supersymmetrized.
We have shown that the introduction of the new singlet fermion leads
to a new contribution to lepton asymmetry and it can be enhanced for certain range of
parameters.
We have also examined how both
the light neutrino mass spectrum and relatively light sterile neutrinos of order a few 100 MeV
can be achieved without being in conflict with the constraints on the mixing between the active and
sterile neutrinos.
One of the noticeable features of our seesaw model is that
the effects of the heavy right-handed neutrinos are cancelled
in seesaw for the tiny neutrino masses, whereas they play an important role in generating
light sterile neutrinos.

\newpage

\noindent {\bf Acknowledgement:}

\noindent SKK is supported in part by the KOSEF SRC program through CQUeST with Grant No.
R11-2005-021 and in part by  KOSEF Grant No. R01-2003-000-10229-0.
CSK is supported in part by CHEP-SRC and in part
by the Korea Research Foundation Grant funded by the Korean Government (MOEHRD)
No. KRF-2005-070-C00030.
\\

\end{document}